\documentstyle[12pt]{article}
\begin{document}
\centerline{\Large\bf Lorentz Gauge Quantization in a}
\centerline{\Large\bf Cosmological Space-time }
\vskip .7in
\centerline{Dan N. Vollick}
\vskip .2in
\centerline{Irving K. Barber School of Arts and Sciences}
\centerline{University of British Columbia Okanagan}
\centerline{3333 University Way}
\centerline{Kelowna, B.C.}
\centerline{Canada}
\centerline{V1V 1V7}
\vskip 0.5in
\centerline{\bf\large Abstract}
\vskip 0.5in
\noindent

It has recently been shown that it is not possible to impose the
Lorentz gauge condition in a cosmological space-time using the
Gutpa-Bleuler method of quantization. It was also shown that it is possible
to add $\nabla_{\mu}A^{\mu}$ as a new degree of freedom to the electromagnetic field and
that this new degree of freedom might be the dark energy which is producing the accelerated
expansion of the Universe.

In this paper I show that
the Lorentz gauge condition can be imposed using Dirac's method of quantizing
constrained dynamical systems. I also compute the vacuum expectation value of the energy-momentum tensor and show that it vanishes.
Thus, in Dirac's approach, the electromagnetic field does not make a contribution to the dark energy.
\newpage
\section{Introduction}
It has recently been shown \cite{Ji1} that it is not possible to maintain
the Lorentz gauge condition using the Gupta-Bleuler approach in a cosmological
space-time. In the past, when the universe is Minkowski, the Gupta-Bleuler
condition $\partial_{\mu}A^{\mu (+)}_{\: in}|\Psi>=0$ is imposed, which
involves only
the positive frequency part of the Lorentz gauge condition. The Universe
then evolves into an out Minkowski region. The transverse
in-positive
frequency modes evolve into
out-positive frequency modes, but the
time-like and longitudinal
in-positive frequency modes evolve into
linear combinations of the
out-positive and out-negative frequency
modes. This produces a violation of the Lorentz gauge condition
$\partial_{\mu}A^{\mu (+)}_{\: out}|\Psi>=0$ in the out region.

The authors then considered quantizing the electromagnetic field in a cosmological
space-time without imposing the Lorentz gauge condition using the action
\begin{equation}
S=\int\left[-\frac{1}{4}F^{\mu\nu}F_{\mu\nu}+\frac{1}{2}\xi(\nabla_{\mu}A^{\mu})^2
\right]\sqrt{g}d^4x
\label{Lagem}
\end{equation}
This gives the
electromagnetic filed an additional scalar degree of freedom, namely $\nabla_{\mu}
A^{\mu}$. They show that super-Hubble modes of this scalar contribute a cosmological constant to
the energy-momentum tensor. This raises the interesting possibility that this additional degree
of freedom may be the dark energy which is producing the accelerated expansion of the Universe.

In this paper I use Dirac's approach \cite{Dir1,Dir2} to quantize the
electromagnetic field in the Lorentz gauge in a cosmological space-time.
Consistency of the time evolution of the Lorentz
gauge constraint introduces a second constraint $\vec{\nabla}\cdot
\vec{E}\approx 0$.
These constraints are first class and
in this approach the full constraints, not just the
positive frequency parts, are imposed on the state vector. Dirac's
approach is shown to give a consistent way to impose the Lorentz gauge condition.

I also compute the vacuum expectation value of the energy-momentum tensor and show that it vanishes.
Thus, in Dirac's approach, the electromagnetic field does not make a contribution to the dark energy.

\section{Dirac Quantization in the Lorentz Gauge}
Consider a Friedman-Robertson-Walker space-time with a metric given by
\begin{equation}
ds^2=a^2(t)\left[-dt^2+dx^2+dy^2+dz^2\right]\; .
\end{equation}
Since Maxwell's equations are conformally invariant in four space-time dimensions
the source free field equations are given by
\begin{equation}
\Box A^{\mu}-\partial^{\mu}\left(\partial_{\alpha}A^{\alpha}\right)=0\; ,
\end{equation}
where $A^{\mu}=\eta^{\mu\nu}A_{\nu}$ and $\Box =\eta^{\mu\nu}
\partial_{\mu}\partial_{\nu}$. The Lorentz gauge condition
\begin{equation}
\nabla_{\mu}\left( g^{\mu\nu}A_{\nu}\right)=0
\end{equation}
is not conformally invariant and can be written as
\begin{equation}
\partial_{\mu}A^{\mu}+2\psi A^{t}=0\; ,
\end{equation}
where
\begin{equation}
\psi=\frac{1}{a}\frac{da}{dt}\; .
\end{equation}
Substituting this into the field equations gives
\begin{equation}
\Box A^{\mu}+2\partial^{\mu}\left(\psi A^{t}\right)=0\; .
\label{FE}
\end{equation}
This equation, together with the Lorentz gauge condition, constitutes the
field equations of the theory.

To quantize this theory I will follow Dirac's method \cite{Dir1,Dir2} of
quantizing the electromagnetic field in Minkowski space-time in the Lorentz gauge.
First the classical theory will be developed and then it will be quantized.

A Lagrangian density which produces ($\ref{FE}$), up to terms that
vanish when the Lorentz gauge condition is imposed, is
\begin{equation}
L=-\frac{1}{2}\left(\partial_{\mu}A_{\nu}\right)\left(\partial^{\mu}A^{\nu}\right)
-2\psi A^{\mu}\partial_{\mu}A_{t}+2\left(\dot{\psi}-\psi^2\right)A_{t}^2\;,
\label{Lag}
\end{equation}
with the action given by $S=\int L\,d^{\,3}x$.
The field equations that follow from this Lagrangian density are
\begin{equation}
\Box A^{\mu}+2\partial^{\mu}\left(\psi A^{t}\right)+2\psi\delta^{\mu}_t
\left(\partial_{\alpha}A^{\alpha}+2\psi A^{t}\right)=0\; ,
\end{equation}
which are equivalent to Maxwell's equations once the Lorentz gauge
condition is imposed.

The canonical momenta densities are given by
\begin{equation}
\Pi^{\mu}=\dot{A}^{\mu}+2\psi A_{t}\delta^{\mu}_{t}\; ,
\end{equation}
the Hamiltonian density is given by
\begin{equation}
h=\frac{1}{2}\Pi^{\mu}\Pi_{\mu}+\frac{1}{2}\left(\partial_kA_{\mu}\right)
\left(\partial^kA^{\mu}\right)+2\psi\Pi^tA_t+2\psi A^k\partial_kA_t
-2\dot{\psi}A_t^2
\label{Ham}
\end{equation}
and the Lorentz gauge condition is given by
\begin{equation}
\chi_1=\Pi^t+\partial_mA^m-4\psi A_t=0\; .
\end{equation}
For consistency it is necessary that
\begin{equation}
\dot{\chi_1}=\{\chi_1,H\}+\frac{\partial\chi_1}{\partial t}=\partial_m\left(
\partial^mA^t+\Pi^m\right)+2\psi\chi_1\approx 0\; ,
\end{equation}
where $\{\;\;\}$ denotes the Poisson bracket, $H=\int hd^{\,3}x$
and $\approx$ denotes a weak equality (i.e. an equality
on the constraint hypersurface).
Thus, there is an additional constraint
\begin{equation}
\chi_2=\partial_m\left(\partial^mA^t+\Pi^m\right)\approx 0\;.
\end{equation}
It is interesting to note that this
new secondary constraint can also be written as
$\chi_2=-\vec{\nabla}\cdot\vec{E}\approx 0$.
The consistency condition $\dot{\chi_2}=\nabla^2\chi_1\approx 0$ does not produce any new constraints.
It is easy to show that $\{\chi_1,\chi_2\}= 0$, so that $\chi_1$ and
$\chi_2$ are first class constraints.

To quantize the theory, in the Schrodinger picture, the dynamical variables $A_{\mu}$ and $\Pi^{\mu}$ become time independent operators satisfying
\begin{equation}
[A_{\mu}(\vec{x}),A_{\nu}(\vec{y})]=[\Pi^{\mu}(\vec{x}),\Pi^{\nu}(\vec{y})]=0
\end{equation}
and
\begin{equation}
[A_{\mu}(\vec{x}),\Pi^{\nu}(\vec{y})]=i\,
\delta^{\nu}_{\mu}\,\delta^3(\vec{x}-\vec{y})\; ,
\end{equation}
where $[\;\;]$ denotes the commutator and I have set $\hbar=1$.
A state vector is introduced that
satisfies the Schrodinger equation
\begin{equation}
i\frac{d}{dt}|\Psi>=H|\Psi>\; ,
\end{equation}
where $H=\int h\,d^{\,3}x$ and $h$
is given by ($\ref{Ham}$). In the standard approach
$2\psi\Pi^tA_t$ would be replaced by  $\psi(\Pi^tA_t+A_t\Pi^t)$.
However, it will be shown that this replacement leads to a divergence
in the Hamiltonian when acting on physical states (defined below).
This replacement will therefore not be done here. The resulting
Hamiltonian will not be Hermitian, but the non-Hermitian
part will vanish when acting on physical states.

The constraints are imposed on the wave function as follows:
\begin{equation}
\chi_1\,|\Psi>=0\;\;\;\;\;\; and\;\;\;\;\;\ \chi_2|\,\Psi>=0\; .
\end{equation}
States that satisfy these constraints are said to be physical states.
The space of physical states will be denoted by $\mathcal{H}$$_{phys}$.
For a consistent quantum theory we require that $[\chi_1,\chi_2]=
\alpha\chi_1+\beta\chi_2$ where $\alpha$ and $\beta$ are operators that
appear to the left of the constraints. This is satisfied since $[\chi_1,\chi_2]=0$.

To preserve the constraints under time evolution it is necessary that
\begin{equation}
\frac{d}{dt}\left[\chi_k|\Psi>\right]=\left\{\frac{\partial\chi_k}{
\partial t}-i\left[\chi_k,H\right]\right\}|\Psi>=0\;.
\end{equation}
Thus, we require that
\begin{equation}
\frac{\partial\chi_k}{\partial t}-i\left[\chi_k,H\right]\approx 0\;,
\end{equation}
where, in the quantum theory, $A\approx 0$ implies that
$A|\Psi>=0$. For $\chi_1$ we have
\begin{equation}
\frac{\partial\chi_1}{\partial t}-i[\chi_1,H]=\chi_2+2\psi\chi_1\approx 0
\end{equation}
and for $\chi_2$ we have
\begin{equation}
\frac{\partial\chi_2}{\partial t}-i[\chi_2,H]=\nabla^2\chi_1\approx 0\; .
\end{equation}
Thus, the constraints are preserved under time evolution.

The constraint $\chi_2$ can be simplified. The term $\partial_m\Pi^m$ involves
only the longitudinal part of $\Pi^m$ and this longitudinal part can be written
as the gradient of a scalar $U$. Thus, $\partial_m\Pi^m=\nabla^2U$. The constraint
$\chi_2$ can therefore be written as
\begin{equation}
\chi_2=\nabla^2(A^t+U)\; .
\end{equation}
Now, $\nabla^2(A^t+U)\approx 0$ over all space has the unique solution $A^t+U
\approx 0$, if the fields vanish at infinity.

The Hamiltonian can be decomposed into transverse and longitudinal/time-like
parts: (see Dirac \cite{Dir1} for the details of the calculation in Minkowski space-time)
\begin{equation}
H_T=\frac{1}{2}\int\left[\Pi^m_{(T)}\Pi^{(T)}_m+\left(\partial_sA^{(T)}_m\right)
\left(\partial^sA^m_{(T)}\right)\right]d^{\,3}x\; ,
\end{equation}
\begin{equation}
H_{L}^{(1)}=\frac{1}{2}\int\left[\partial_r(U-A^t)\partial^r(U+A^t)+(\partial_mA^m
-\Pi^t)(\partial_mA^m+\Pi^t-4\psi A_t)\right]d^{\,3}x
\end{equation}
and
\begin{equation}
H_{L}^{(2)}=-2\dot{\psi}\int A_t^2 d^{\,3}x \; .
\end{equation}
Note that $H_{L}^{(1)}\approx 0$. In flat space-time, with $\psi=0$,
$H_{L}^{(2)}=0$, so that
the longitudinal and time-like degrees of freedom do not contribute to
the field energy. This is not the case in a cosmological space-time with
$\dot{\psi}\neq 0$. The transverse part of the Hamiltonian is the same
as it is in Minkowski space-time.

It is interesting to note that if $\,2\psi\Pi^tA_t\,$ were replaced by
$\,\psi(\Pi^tA_t+A_t\Pi^t)\,$ there would be an additional term
$\psi[A_t(x),\Pi^t(x)]\sim i\psi
\delta^3(0)$ in $H_{L}^{(2)}$, so that when the universe is
expanding (or contracting) the evolution of $|\Psi>$ would not be well
defined. It is also important to note that although $H$ is not
Hermitian it satisfies
\begin{equation}
H\approx\frac{1}{2}\int\left[\Pi^m_{(T)}\Pi^{(T)}_m+\left(\partial_sA^{(T)}_m\right)
\left(\partial^sA^m_{(T)}\right)\right]-4\dot{\psi}A_t^2]d^{\,3}x\; .
\end{equation}
Thus, when acting on physical states $H$ is Hermitian.

To simplify the constraints and dynamics consider the ket
\begin{equation}
|\Psi_M>=exp\left[-2i\psi\int A_t^2d^{\, 3}x\right]|\Psi>\; .
\end{equation}
The constraints satisfied by $|\Psi_M>$ are
\begin{equation}
\left(\Pi^t+\partial_mA^m\right)|\Psi_M>=0\;\;\;\;\;\;\; and \;\;\;\;\;\;\;\;
\left(\partial_m\Pi^m+\nabla^2A^t\right)|\Psi_M>=0\;
\end{equation}
and $|\Psi_M>$ satisfies the equations of motion
\begin{equation}
i\frac{d}{dt}|\Psi_M>=H_T|\Psi_M>\;.
\end{equation}
This ket therefore satisfies the Minkowski space constraints and equation
of motion. Thus, we see that $|\Psi>$ is equal to a Minkowski state
vector multiplied by an operator valued phase factor. It is interesting to
note that $<\Psi|\Psi>=<\Psi_M|\Psi_M>$ and that
\begin{equation}
<\Psi|\Omega\left(\Pi^{m},A_{\mu}\right)|\Psi>=
<\Psi_M|\Omega\left(\Pi^{m},A_{\mu}\right)|\Psi_M>\; ,
\end{equation}
where $\Omega$ is any operator that depends on $\Pi^m$ and $A_{\mu}$.

An Operator $\Omega$ is said to be physical if $\Omega|\Psi>\in\mathcal{H}$$_{phys}$
for all $|\Psi>\in\mathcal{H}$$_{phys}$. Physical operators must
then satisfy $[\Omega,\chi_k]=0$. Since the constraints involve
only time-like and longitudinal components any operator that depends
only on the transverse components will be physical. The electromagnetic
field $F_{\mu\nu}$ is also physical, but the potentials $A_{\mu}$ are not.

As is well known there exists a residual gauge freedom of the form $\bar{A}^{\mu}=
A^{\mu}+\nabla^{\mu}\chi$ with $\Box\chi=0$ even once the Lorentz gauge is imposed.
It is easy to show that $H_T$ and $H_L^{(1)}$ are invariant under this
transformation while $H_L^{(2)}$ is not. From (27) one can see that a
gauge transformation of this type only changes the phase of the state vector.

To set up a Fock space representation the operators
\begin{equation}
a^{\mu}_{\vec{k}}=\int e^{-i\vec{k}\cdot\vec{x}}\left[kA^{\mu}(\vec{x})+i
\Pi^{\mu}(\vec{x})\right] d^{\, 3}x
\end{equation}
and
\begin{equation}
a^{\dag\,\mu}_{\vec{k}}=\int e^{i\vec{k}\cdot\vec{x}}\left[kA^{\mu}(\vec{x})-
i\Pi^{\mu}(\vec{x})\right]d^{\, 3}x
\end{equation}
can be introduced, where $k=|\vec{k}|$. These operators satisfy the
standard commutation relations
\begin{equation}
\left[a^{\mu}_{\vec{k}}\,,\,a^{\nu}_{\vec{k}'}\right]=\left[a^{\dag\,\mu}_{\vec{k}}\,,
\,a^{\dag\,\nu}_{\vec{k}'}\right]=0
\end{equation}
and
\begin{equation}
\left[a^{\mu}_{\vec{k}}\,,\,a^{\dag\,\nu}_{\vec{k}'}\right]=(2\pi)^3(2k)\eta^{\mu\nu}
\delta^3(\vec{k}-\vec{k}')\;.
\end{equation}
The spatial parts of these operators can also
be decomposed into transverse and longitudinal parts.

A vacuum state $|0>$ can also be introduced that satisfies
\begin{equation}
\left(\Pi^t+\partial_mA^m-4\psi A_t\right)|0>=0,\;\;\;\;\;\;\;\;\;\;\;
\left(\partial_m\Pi^m+\nabla^2A^t\right)|0>=0
\label{vac1}
\end{equation}
and
\begin{equation}
a^{(T)}_{m\vec{k}}\,|0>=0\; .
\label{vac2}
\end{equation}
These conditions are consistent because $a^{(T)}_{m\vec{k}}$ commutes
with the two constraints.
The operators $a^{(T)}_{m\vec{k}}$ act as annihilation operators and the
operators $a^{\dag (T)}_{m\vec{k}}$ act as creation operators.

If $H_T$ is normal ordered so that $H_T|0>=H_T|0_M>=0$ then
\begin{equation}
|0(t)>=exp\left[2i\psi(t)\int A_t^2d^{\, 3}x\right]|0_M>\;,
\end{equation}
where $|0_M>$ is the standard Minkowski vacuum. Now consider
a cosmological space-time with asymptotic \textit{in} and \textit{out}
Minkowski regions.
If the initial
state of the field, in the past asymptotic region, is $|0_M>$ the final
state of the field, in the future asymptotic region, is also $|0_M>$.
There is, therefore, no particle production for the electromagnetic field
in a cosmological space-time.  This is in agreement with results
obtained earlier by Parker \cite{Par1,Par2}.

\section{Dark Energy}
Since the field equations of the theory are Maxwell's equations in the Lorentz gauge the energy-momentum tensor will
be given by the standard expression
\begin{equation}
T_{\mu\nu}=F_{\mu\alpha}F_{\nu}^{\;\;\alpha}-\frac{1}{4}g_{\mu\nu}F^{\alpha\beta}F_{\alpha\beta}\; .
\end{equation}
Now consider $E=-\int T^t_{\;\; t}d^3x$. A short calculation shows that
\begin{equation}
E\approx \frac{1}{a^4}\left[H+2\dot{\psi}\int A_t^2\;d^3x+E_s\right]\;,
\end{equation}
where $E_s$ is a surface term. This surface term will not contribute to the
average energy density it we consider an infinite volume. Since the energy density
must be spatially uniform in an FRW space-time the surface term may be dropped. Note
that $E$ is invariant under the the action of the residual gauge transformations.

Thus, in the ground state, where $<0|H|0>=-2\dot{\psi}\int A_t^2\,d^3x$, we find that $<0|E|0>\approx 0$.
This implies that
\begin{equation}
<0|\rho|0>\approx 0\;.
\end{equation}
The vacuum expectation value of the energy density therefore vanishes. The pressure follows from the continuity equation
(in conformal coordinates)
\begin{equation}
\frac{d}{dt}(a^2\rho)+a^2\psi(\rho+3P)=0\;.
\end{equation}
Thus, when $\psi\neq 0$ the vacuum pressure must vanish.

We therefore conclude that the vacuum expectation value of the energy-momentum tensor vanishes when the field
is in the ground state. Thus, there is no dark energy when the electromagnetic field is quantized
in the Lorentz gauge using Dirac's procedure.

\section{Conclusion}
It has recently been shown \cite{Ji1} that it is not possible to maintain
the Lorentz gauge condition in a cosmological space-time using the
Gupta-Bleuler approach. The authors also considered quantizing the electromagnetic field using the
action (\ref{Lagem}) without imposing the Lorentz gauge condition. This introduces a new scalar degree of freedom,
$\nabla_{\mu}A^{\mu}$, and they show that super-Hubble modes of this scalar contribute a cosmological constant to the
energy-momentum tensor. This raises the interesting possibility that this new degree of freedom may be the dark energy which
is producing an accelerated expansion of the Universe.

In this paper I quantized the electromagnetic field in
the Lorentz gauge in a cosmological space-time using Dirac's method for
constrained dynamical systems. Consistency of the Lorentz gauge condition under time evolution introduces a second constraint $\nabla\cdot\vec{E}\approx 0$. These two constraints are maintained under time evolution, showing that it is possible to impose the Lorentz gauge condition in a cosmological space-time.
I also showed that there is no particle production in cosmological space-times, which is consistent with earlier results obtained by Parker \cite{Par1,Par2}.

I also computed the vacuum expectation value of the energy-momentum tensor and showed that it vanishes.
Thus, in Dirac's approach, the electromagnetic field does not make a contribution to the dark energy.

\section*{Acknowledgements}
This research was supported by the  Natural Sciences and Engineering Research
Council of Canada.

\end{document}